\documentclass[11pt]{article}
\usepackage{apalike,setspace,ifthen,endfig,graphics}

\graphicspath{{./Figures/}}

\newcommand{\DRAFT}{1=0}

\newcommand{\ps}[3]{\resizebox{#1}{#2}{\includegraphics{#3}}}

\newcommand{\dgr}{ {^\circ} }

\begin{document}


\pagestyle{plain}
\begin{singlespace}

\vspace*{-1.0in}
{ \ifthenelse{\DRAFT}{\small}{} \flushright
  Submitted to the {\em Journal of Neuroscience}, \hspace*{-.5in} \\
  Behavioral/Systems Neuroscience Section, \hspace*{-.5in} \\
  Dr. Stephen G. Lisberger, Section Editor \hspace*{-.5in} \\ }
\ifthenelse{\DRAFT}{}{\vskip50pt}

\ifthenelse{\DRAFT}{
  \vspace*{.25in} \noindent {\Large \sf Draft: Please Do Not Distribute}
  \vskip24pt
}{}

\begin{center}
  {\huge \bf 
Temporal structure in neuronal activity
during working memory in Macaque parietal cortex
    } \\[12pt]
	\vskip12pt
  {\Large Bijan Pesaran$^{1}$, John Pezaris$^{2}$, Maneesh Sahani$^{2}$, 
	Partha P.\ Mitra$^{3}$ and Richard A.\ Andersen$^{2,4}$} \\[24pt]
	\vskip12pt
  {\large
    \begin{minipage}{2.5in}{ \begin{center}
          $^{1}$ Division of Physics, Mathematics and Astronomy \\
          California Institute of Technology \\
          Pasadena, CA\, 91125 \end{center} }   
    \end{minipage}
    \begin{minipage}{2.5in}{ \begin{center}
          $^{2}$ Computation and Neural Systems Program \\
          California Institute of Technology \\
          Pasadena, CA\, 91125 \end{center} }
    \end{minipage}
	\vskip24pt
    \begin{minipage}{2.5in}{ \begin{center}
          $^{3}$ Bell Laboratories \\
          Lucent Technologies \\
          Murray Hill, NJ\, 07974 \end{center} }
    \end{minipage}
    \begin{minipage}{2.5in}{ \begin{center}
          $^{4}$ Division of Biology \\
          California Institute of Technology \\
          Pasadena, CA\, 91125 \end{center} }
    \end{minipage}}
\end{center}
\vskip12pt

{ \flushleft
\noindent 
\hspace*{-3mm} \begin{tabular}{ll}
  {\bf Abbreviated Title:}  & Temporal structure during working memory  \\ 
  {\bf Page Count:}         &  41 (excluding figures) \\
  {\bf Figure Count:}       &  15 \\
  {\bf Table Count:}        & 0 \\
  {\bf Word Counts:}        & Abstract, 241;\, Introduction, 519;\, Discussion, 1550
\end{tabular} }
\vskip8pt

\ifthenelse{\DRAFT}{\vskip24pt}{\hspace*{-.75in}} { \noindent {\bf
    Acknowledgments:} This work was supported by NIH grant EY05522-21, 
        ONR grant N00014-94-0412, 
        Bell Labs, Lucent Technologies internal funding, 
        the Keck Foundation, the Sloan Foundation, 
        and the Workshop for the Analysis of Neural Data,
        MBL, Woods Hole, MA 
        ({\tt http://www.vis.caltech.edu/$\sim$WAND})} 

\vskip12pt

\ifthenelse{\DRAFT}{}{\hspace*{-.75in}}
{ \noindent 
  {\bf Please address correspondence to:} \\
  Richard A. Andersen \\
  Division of Biology 216-76 \\
  California Institute of Technology \\
  Pasadena, CA\, 91125 \\
  {\tt andersen@vis.caltech.edu} \\
  (626) 395-8336 (Office) (626) 795-2397 (Fax)  }

\end{singlespace}
\doublespacing
\newpage
\pagestyle{myheadings} 
\markboth{Pesaran et. al.}{Pesaran et. al.}

\centerline{\Large \bf Abstract}
\vskip12pt

A number of cortical structures are reported to have elevated single
unit firing rates sustained throughout the memory period of a working
memory task.  How the nervous system forms and maintains these
memories is unknown but  reverberating neuronal network activity is
thought to be important.  We studied the temporal structure of single
unit (SU) activity and simultaneously recorded local field potential
(LFP) activity from area LIP in the inferior parietal lobe of two
awake macaques during  a memory-saccade task.   Using multitaper
techniques for spectral analysis, which play an important role in
obtaining the present results, we find elevations in spectral power in
a 50--90Hz  (gamma) frequency band during the memory period in both
SU and LFP activity.  The activity is tuned to the direction of the
saccade providing evidence for temporal structure that codes for movement
plans during working memory.
We also find SU and LFP activity are coherent during the memory period
in the 50--90Hz gamma band and no consistent relation  is
present during simple fixation.  
Finally, we find organized LFP activity
in a 15--25Hz frequency band that may be related to movement
execution and preparatory aspects of the task.   
Neuronal activity could be used to
control a neural prosthesis but
SU activity can be hard to isolate with cortical implants.
As the LFP is easier to acquire than SU activity, 
our finding of rich temporal structure
in LFP activity related to movement planning and execution may
accelerate the development of this medical application.

\vskip 24pt


\newpage

Working memory is a brain system requiring the temporary storage and
manipulation of information necessary for the performance of complex
cognitive tasks \cite{Baddeley1992}.  The neurophysiological basis of
working memory is studied in non-human primates by recording neural
activity during delayed-response tasks  \cite{Fuster1995}.
Cue-selective elevated single unit firing rates have been recorded
during the delay period in many brain areas during different versions
of the task \cite{FusterJervey1982,BruceGoldberg1985,GnadtAndersen1988,MiyashitaChang1988,FunahashiBruceGoldman-Rakic1989,KochFuster1989,MillerEricksonDesimone1996,ZhouFuster1996}.  How this
neural activity is sustained is unknown but may be important to
understanding the neural basis of working memory
\cite{Goldman-Rakic1995}.  Converging evidence points to the
importance of a distributed recurrent neuronal network
\cite{Goldman-Rakic1988} and  reverberating network activity has long
been suggested as a  possible mechanism for short-term memory
\cite{LorentedeNo1938,Hebb1949,Amit1995,Seung1996,Wang1999}.

Measures with the potential to capture correlated neural activity   on a
millisecond time scale may be needed to resolve reverberating memory
activity.   The dynamical structure of neuronal activity has been the
source of much interest as a temporal code (for a review see Singer
and Gray (1995)
\nocite{SingerGray1995}) however much of this work emphasizes
stimulus-induced  activity and its relation to perception
\cite{EckhornBauerReitboeck1988,GrayKonigSinger1989,EngelKonigSinger1990,deCharmsMerzenich1996,BorstTheunissen1999}.  
Other work in somatosensory and motor
areas reports temporal structure in activity during a period before
action  \cite{SanesDonoghue1993,BresslerCoppolaNakamura1993,MurthyFetz1996a,MurthyFetz1996b,RoelfsemaEngelSinger1997} and electroencephalography (EEG)
studies in humans  report sustained oscillatory  responses during
working  memory \cite{Tallon-baudryKreiterBertrand1999}.

Elevated single unit (SU) activity in parietal 
cortex during a memory-saccade task
was first reported on the lateral bank of the intraparietal sulcus
(area LIP) \cite{GnadtAndersen1988} and this area has now been
implicated in a variety of cognitive processes  (for a review see
Andersen (1995)
\nocite{Andersen1995}).   More recent work reports similar memory
activity in parietal cortex before reaches
\cite{SnyderBatistaAndersen1997} and grasps
\cite{SakataTairaMurata1995} relating parietal cortex to movement
plans in general.  There is great interest in helping paralyzed
patients by using cortical SU activity to control a prosthesis
\cite{KennedyBakay1998,ChapinMoxonNicolelis1999} 
and the relation of parietal SU activity to
movement planning makes it a potentially useful signal
\cite{ShenoyKureshiAndersen1999}.  However, the difficulty of
isolating  SU activity with cortical implants has presented  a major
hurdle to the development of a neural prosthesis.    The local field
potential (LFP), which
measures activity in a population of neurons, is easier to acquire
than SU activity.  Consequently the  presence of temporal
structure in LFP activity  could prove important for 
the development of this application.

We recorded SU and LFP activity from two macaques during a
memory-saccade task  using a single tetrode located in area LIP.  SU
activity has been previously examined in area LIP during this task
\cite{GnadtAndersen1988,ChafeeGoldman-Rakic1998}  and previous
analysis of this data has investigated SU activity for temporal
structure.  \cite{PezarisSahaniAndersen1997a} looked for oscillations
using  autocorrelation functions but averaged over long periods of
time and reported their absence.  \cite{PezarisSahaniAndersen1999}
showed evidence for structure in  auto- and cross-covariations
between spike trains without examining significance.   LFP activity 
in parietal cortex has not been previously examined.  In this
work, we use modern techniques of multitaper spectral analysis to
investigate temporal structure in SU and LFP activity.  We
find significant structure in 
the spectrum of SU and LFP activity and the coherency between
them.  We also find additional significant spectral
structure in LFP activity 
that codes for movement execution and preparatory aspects of the task.
These results provide evidence for
memory fields of temporal structure (dynamical memory fields) that may
reflect the columnar organization of cortex.

\section*{MATERIALS AND METHODS}

\subsection*{Animal preparation}  Recordings were made from two adult 
male Rhesus monkeys ({\it Maccaca mulatta}).  Animals were 
fitted with a stainless steel head post
embedded in a dental acrylic head cap to fix their head position,  a
scleral search coil to record eye position and a stainless steel
recording  chamber was placed over a craniotomy to gain access to the
cortex.  All surgical procedures and animal care protocols were
approved by the California Institute of Technology Institutional
Animal Care and Use Committee and were in accordance with National
Institutes of Health Guidelines.

\subsection*{Behavioral task} Recordings were made while animals performed
memory-saccade task (see Fig. 1).  Each trial of this task begins with the
illumination of a central fixation light, to which the animal saccaded.  As
long as the fixation light was present, the animal was required to maintain
fixation within a $2\dgr$ circular window.  After a post-foveation
background period of $1-2$ seconds, a target light was flashed for $100ms$ at
one of 8 fixed stimulus locations evenly distributed on a $10\dgr$ circle.
Following the target flash, the monkey had to maintain fixation for a
further period of $1000ms$, at the end of which, the fixation light was
extinguished, and the animal required to saccade to the remembered location
of the flashed stimulus.  For accurate saccades, the target was
reilluminated for a minimum of $500ms$, often triggering a corrective
saccade, and the animal required to fixate at the new location while the
target remained on.  At the completion of a successful trial, the animal
was rewarded with a drop of water or juice.  Target locations were randomly
interleaved to collect between $10-15$ successful trials for each location
in blocked fashion.

\subsection*{Electrophysiological recordings} Electrical activity was
recorded from the behaving monkey using single tetrodes
\cite{ReeceO'Keefe1989} adapted for use in the awake monkey preparation
\cite{PezarisSahaniAndersen1997a}.  This is extensively described elsewhere
\cite{Pezaris2000}, and is briefly summarized here.  Tetrodes were
placed
in a fine guide tube and and positioned using a standard hydraulic
microdrive (Fred Haer Corp, Brunswick, ME).  Neural signals were amplified
by a custom four-channel headstage amplifier ($A=100$,
feeding a custom four-channel variable-gain preamplifier ($A=1-5000$,
nominally set to 200, \cite{Pezaris2000}) and anti-alias
filters (9-pole elliptical low-pass, $f_c=10$~kHz, Tucker-Davis
Technologies (TDT), Gainsville, FL) before being digitized with a
four-channel instrumentation-grade 16-bit analog to digital converter
($f_s=20$~kHz, also TDT).  Digital data were then streamed to disk and
written to CD-ROM.  The polarity of the LFP was reversed to give
positive-going spike activity.  Continuous extracellular traces were processed
off-line to extract and classify spike events and calculate the LFP.
Figure 2 presents the  extracellular potential from a channel of the
tetrode.  Panel a) presents activity during one trial and panel b)
presents activity during the memory period on an expanded time base.

\subsection*{Spike Sorting}
SU activity was extracted from the digitized recordings by an
automated procedure which identified and sorted spike waveforms into
clusters, each presumed to arise from a single cell.  The algorithm has
been motivated and described by Sahani et al. (1998) 
\nocite{SahaniPezarisAndersen1998} and in detail by
Sahani (1999) and will be summarized here.

Prior to spike sorting, the recordings were bandpass filtered (0.6 - 6
kHz).  A statistical model describing the distribution of spike shapes
was then fit to waveforms extracted from 30s of data, as described
below.  This model was used to classify spike events in the rest of
the recording.

First, candidate spike times were identified by comparing the signal
to a threshold of 3 times the root-mean-square (RMS) signal
value on each channel.  Spikes were accepted when the trace i) crossed
and remained above the threshold for at least 0.1ms on at least one
channel; ii) crossed the threshold on the other channels either within
0.1ms of this time, or else not within 1ms; iii) did not remain above
threshold for longer than 1ms; and iv) did not cross the threshold
again within 1ms.  These constraints reduced the number of overlapped
spike events in order to reduce bias in estimating the spike waveform
model.

A 2.4ms segment (48 samples per channel) of data was then extracted
from all four channels, centered on each identified spike time.  A (2
$\cdot$ RMS) thresholded center of mass was calculated for each spike
waveform, and the segment resampled by interpolation to yield 24
samples per channel (1.2ms), with the center of mass falling
one-quarter of the way into the waveform.  The precision of this
center of mass alignment was 4-8 times the original sampling
frequency.  The different channels were then concatenated to yield
96-dimensional event vectors.

The background noise covariance expected in these event vectors was
estimated using 1.2ms segments extracted from the recording at times
when no threshold-crossing was seen.  The event vectors were then
transformed to lie in a space where this noise covariance was
whitened, and a mixture of a single Gaussian and a uniform density fit
to them.  The principal eigenvectors of the covariance of the Gaussian
provided robust estimates of the principal components of the data in
the noise-whitened space.  Low-dimensional event vectors were obtained
by projecting each transformed event onto the four leading
eigenvectors of the Gaussian covariance.

Events were clustered by fitting a mixture of Gaussian distributions
to the low-dimensional event vectors.  The covariance of each Gaussian
was fixed at the identity matrix (since the background noise is white
in the transformed event space).  A uniform component was introduced
in the mixture to reduce the effect of outliers.  The fit was
performed using the Relaxation Expectation-Maximization algorithm
(Sahani, 1999; see also Ueda and Nakano 1994) \nocite{UedaNakano1994}
which helped to avoid
problems of local minima.  The correct number of Gaussian components
was determined by cascading model selection \cite{Sahani1999}.

Each Gaussian component in the mixture was taken to represent the
distribution of spikes expected from a single cell, and events were
assigned to cells according to a maximum \textit{a posteriori} rule.
The autocorrelogram of spikes assigned to each cluster was checked to
ensure that no violations of the refractory period were seen.  To
ensure that the clustering was robustly determined, the segment of
data used to fit the model was varied and only models that were
consistent for all segments were included in database.

\subsection*{Data analysis}

The data are SU activity (a point process) and the LFP (a
continuous valued time series) making this a hybrid data set.
Spectral analysis provides a unified framework for the
characterization of hybrid processes and we use multitaper methods of
spectral estimation developed in Thomson (1982)
\nocite{Thomson1982}  to construct
estimators for all spectral quantities (see {\bf Appendix}).   
These estimates are
presented and evaluated in \cite{PercivalWalden1993} and are applied
to a wide range of neurobiological data in \cite{MitraPesaran1999}.
Correlation function measures such as the auto- and
cross-correlation function characterize the same statistical structure
in time series as spectra and cross-spectra, however, as we discuss
in a later section, spectral estimates offer significant advantages
over their time domain  counterparts.  

Spike events were binned at 1ms time resolution to give spike trains.
LFP time series were calculated from the extracellular recordings from
one tetrode channel by lowpass filtering the signal at 200Hz using
multitaper projection filters \cite{Thomson1994c}.  The out-of-band
suppression properties of these filters ensured minimal corruption of
the LFP estimate by SU activity.  Controls, described below, were performed
that confirm this.

\vskip8pt
\noindent{\sc Mean Responses:}  The mean response of SU activity, the
peri-stimulus time histogram (PSTH),  was calculated by counting the
number of spikes per 1ms bin  and averaging across trials for each
saccade direction aligned to the initial target onset, followed by
smoothing with a gaussian kernel, ($\sigma = 10ms$).  The mean
response of the LFP was calculated  analogously to SU mean responses
by averaging across trials for each saccade direction aligned to the
initial target onset and smoothing with a gaussian kernel,  ($\sigma =
10ms$).  Trials were aligned to the saccade before averaging with no
qualitative change in the results.

\vskip8pt
\noindent{\sc Spectral analysis:} Details of the multitaper techniques
used to estimate the spectral  quantities are  presented in the
{\bf Appendix}.  In brief, spectral analysis is the study of time series in the
frequency domain.  The frequency domain is preferred over the time
domain for multiple reasons, an essential one being that neighboring
points in time are usually highly correlated while neighboring points
in frequency are more nearly independent.  Multitaper methods for
spectral estimation use special data tapers.  A segment of
data is multiplied by a data taper before Fourier transformation.
called prolate spheroidal functions (also known as Slepian functions)
that are maximally concentrated in a bandwidth in frequency, $W$, for a
specified  duration in time, $T$ \cite{SlepianPollack1961}.  In addition
to being maximally concentrated in frequency and time, there are many
Slepian functions for a given choice of $T$ and $W$, and they
are orthogonal to each other.   Specifically, there are $K$ functions
that are concentrated where $K=2p-1$, and $p=TW$, the time-bandwidth
product.  The use of multiple Slepian functions as data
tapers give multitaper estimates their minimum bias and variance
properties.

Three periods during the trial were investigated: {\em baseline}, {\em
memory} and {\em perisaccadic}.  The baseline period was defined as
extending 750ms to 250ms before the initial target onset.  The memory
period was defined as extending 450ms to 950ms following the initial
target onset.  The perisaccadic period was defined as extending from
250ms before the saccade to 250ms after the saccade.  Baseline
activity was estimated  by pooling activity from all successful
trials.  Memory and peri-saccadic activity were estimated by pooling
activity from successful trials according to saccade direction.

Two saccade directions of interest were defined for SU and LFP
activity,  {\em preferred} and {\em anti-preferred}.  The preferred
direction was the direction that elicited the maximum firing rate
during the period from the initial target onset to the offset of the
fixation light.  The anti-preferred direction was defined as the
opposite direction to the preferred direction.   In this study, 
there were no cases
of multiple SUs recorded at the same site with different
preferred directions.  The preferred direction of the LFP was defined
to be the same as the preferred direction of simultaneously recorded
SU activity.  Activity from different locations was aligned to the
preferred direction before estimating population average quantities.

In all cases, time-frequency representations of the activity were
calculated  on a 500ms window that was stepped by 100ms between
estimates  through the trial.  The time index was aligned to the
center of the analysis window.  As a control, window size was varied
from 200--750ms without significant change in the results.  In
contrast to the characterization of baseline, memory and perisaccadic
activity described above, time-frequency representations presented
were estimated by averaging trials aligned to the initial target
onset.  Trials were aligned to the saccade before averaging with no
qualitative change in the results.  Slepian functions with
different time-bandwidth products were used to
estimate SU and LFP spectra, and SU-LFP coherency
that best resolved significant structure
in the data.

\vskip8pt
\noindent{\sc SU and LFP spectra:} The spectrum, $S(f)$, is a real
quantity that is a function of frequency for a given window in time
and is a measure of temporal organization.  For example, if single
unit activity tends to cluster in time with a typical spacing, then
the spectrum will show a peak at the inverse of that spacing.  The
spectrum is the Fourier transform of the auto-correlation function but
the use of data tapers makes estimates of the spectrum less
susceptible to bias and variance problems compared to the
auto-correlation function.

SU spectrum was estimated on a 500ms window using 9 Slepian data
tapers with time-bandwidth product, $2p = 10$ giving a frequency
resolution of $\pm 10$Hz.  The spectrum of SU activity estimated for
each cell giving the {\em single cell} estimate.  SU activity from
each cell was aligned according to preferred direction and  averaged
for each saccade direction across all cells in each monkey to give the
{\em population average}.  SU spectrum significance levels were estimated with
multitaper methods using the jackknife method \cite{ThomsonChave1991}.

The LFP spectrum was estimated on a 500ms window using 5 Slepian
data tapers with time-bandwidth product, $2p = 6$ giving a frequency
resolution of $\pm 6$Hz.  The spectrum of LFP activity was estimated
for each recording giving the {\em single site} estimate.  LFP
activity from each site was aligned according to preferred direction
and  averaged for each saccade direction across all sites in each
monkey to give the {\em population average}.  LFP spectrum significance levels
were estimated with multitaper methods using the jackknife method.

\vskip8pt
\noindent{\sc Coherency:} The coherency, $C(f)$, is a complex quantity
that is a function of  frequency for a given window in time and
measures the degree of predictability of one process using a linear
function of the other  \cite{Brillinger1975,RosenbergAmjadHalliday1989}.  Here we use a hybrid version of
the coherency for point process and ordinary time series  data to
measure the relation between SU and LFP activity.

The magnitude of the coherency lies between zero and one and is called
the coherence.  Coherence indicates the strength of the relationship
between the processes.   The coherency is used in preference to the
cross-correlation function which is usually estimated to measure the
relation between two processes.  In particular, increases in coherence
are not due to  changes in firing rates as the coherency is explicitly
normalized by the spectrum of the  individual processes.  This  is not
true for the cross-correlation function which may be normalized in an
{\it ad-hoc} fashion, but is not normalized by simple means.

SU-LFP coherency was estimated on a 500ms window using 19 Slepian
data tapers with time-bandwidth product, $2p = 20$ giving a frequency
resolution of $\pm 20$Hz.  The coherency of SU and LFP activity was
estimated for each cell giving  the {\em single cell} estimate.
SU-LFP coherency estimates were aligned according to preferred
direction and the complex coherency values averaged to give the {\em
population coherency}.  SU-LFP coherency significance levels were estimated
with multitaper methods using the jackknife method.

\vskip8pt
\noindent{\sc Spike-triggered Potential:} The relation between single
unit activity and the extracellular potential was  assessed with the
use of a spike-triggered average of the raw data during the baseline
period and memory period prior to a saccade in the preferred direction
for that cell.  200ms segments of the potential 
centered on the SU event time were extracted.  
Spike waveforms were suppressed by
subtracting a 2ms mean spike shape waveform.  The traces were then
averaged to give the spike-triggered potential.   The spike-triggered
potential provided a measure of association between SU activity and
the extracellular potential and provided a control against corruptions
in the LFP that might result from incompletely filtering out SU spike
activity.  95\% confidence intervals were calculated by estimating the
standard error of the mean.

\vskip8pt
\noindent{\sc Phase Histogram:}  The LFP was filtered between 50--90Hz
using projection filters \cite{Thomson1994c} to give a complex valued
signal.  The frequency band was selected according to results of the
coherency analysis.
The phase of this
signal was sampled at spike event times during the baseline and memory
periods.  Phases were pooled from the preferred direction during the
memory period, and from all trial conditions during the baseline.  The
phases were then pooled across all cells in the study.  These phases
are the frequency domain version of the spike-triggered potential.
The histogram  of phases between the baseline and memory periods
provided a control for the coherency estimate as estimation of the
histograms was not limited 
to small numbers of spike events from a particular cell.
A one-sample Kolmogorov-Smirnov test (KS test) was used  to determine
significance level for deviations of the phase histogram during the
memory period from a uniform distribution \cite{Rao1965}.  A
two-sample KS test was used to determine the significance level for
deviations of the phase  histogram during the memory period from the
baseline period.

\section*{RESULTS}

The database for this study contained 16 SUs recorded at 16 sites in one
monkey and activity from 24 SUs recorded at 17 sites in another monkey.
Since were were primarily interested in memory period activity, a
subjective evaluation was made during data collection to make recordings
only when tuned memory activity was present.  Subsequently, during off-line
analysis, recordings were further selected for containing at least one
clearly and stably isolated cell, using criteria described in the {\bf Methods}
section above.

\subsection*{Mean SU and LFP responses}

Mean SU activity in area LIP during memory-guided saccades has been
characterized in previous reports \cite{GnadtAndersen1988,BarashBracewellAndersen1991a}.
Our results agreed with those
reports and we found 28 of 40 neurons (70\%) had significant
memory period activity 
($p<0.05$).  Figure 3 shows the response of a typical cell with memory
activity.   The  eight trial conditions are displayed in a spatial map
with eye position and behavioral events shown beneath the PSTH and spike
rasters from each trial shown above.  The preferred direction of this
cell is for saccades to the left.  Activity from this SU is presented
as a running example of SU activity in subsequent single cell figures below.

Figure 4 a) shows the PSTH for a single unit across trial  conditions
in a 2d plot.  Line plots of the spike rate in the preferred and
anti-preferred directions are shown in panel b).  Figure 4 c) shows the
mean LFP response for a single site across trial directions and panel
d) shows line plots in the preferred and anti-preferred directions.
The increase of the LFP with narrow tuning during and following the
initial target illumination is similar to SU activity.  However, the
broad, tuning of the LFP across saccade directions  peri- and
post-saccadically is not observed in SU activity.   This feature is a
result of the increase in the mean LFP response occurring with a
systematic increase in latency across trial conditions.  Following the
saccade, there is  also a suppression in the mean LFP response that is
narrowly tuned.  These features are seen in recordings from all sites
recorded from in area LIP in both monkeys. 
LFP activity recorded from this site is presented
as a running example of LFP activity in subsequent single site figures below.

\subsection*{Temporal structure in SU and LFP activity during working memory}

The maintenance of sustained memory  activity may be due to the
intrinsic dynamics of local or long-range networks \cite{Hebb1949}.
These dynamics may not be revealed by a study of elevated mean firing
rates or mean responses as they could involve the temporal
organization of neuronal activity.  Here we use multitaper techniques
for spectral analysis to study the temporal  structure of SU and LFP
activity.

Figure 5 shows spectrograms and individual spectra of SU activity in
the preferred direction for a single cell and the population average
from one monkey.   Panel a) shows the spectrum of SU activity during
the memory (red) and baseline (blue) periods from a typical single
unit in the solid line.  The high-frequency limit of the spectrum in
each case is shown by the horizontal dotted line.  The spectrum of a poisson
process with the same firing rate would fall on this line, therefore
deviations of the spectrum  from this line are evidence of temporal
structure in SU activity.  
95\% error bars  are shown by thin dashed lines in each
case.   Two significant features are present in the memory period
spectrum that are absent from the baseline period.  This indicates the
presence of temporal structure during working memory.  The first
significant feature  is a suppression in power at frequencies below
20Hz.  The second is a peak in power at a higher frequency band, in
this case  centered at 50Hz.      Panel b) shows a single unit
spectrogram for the same unit shown in panel a).   The spectrogram
presents the spectrum as a function of time and the spectral structure
is sustained throughout the memory period, beginning at the initial
target illumination and continuing through the saccade.

Panels c) and d) of Figure 5 show the spectrum and spectrogram for the
population average.  The significant features of suppression of power
at low frequencies and the elevation in the gamma band during the
memory period are also present and are sustained throughout the  trial
across a population of single units.  Of the 40 single units recorded
for this study, memory period activity from 
21 neurons (53\%) showed significant spectral peaks
($p<0.05$) and 22 neurons (56\%) showed significant spectral suppression
($p<0.05$).   In all cases, SUs that
showed significant spectral structure had elevated mean firing rates
during the memory period compared to the baseline period.  
Inspection of the individual spectra
reveals that spectral suppression out to 20Hz is present across  the
population while the peak frequency  exhibits more variability and can
occur at frequencies as low as 25Hz and as high as 70Hz.  
Structure in SU memory period activity during
saccades to the  anti-preferred direction is difficult to resolve
because of low spike number but, similar to baseline activity, we find
no significant features in the spectrum.

If temporal structure in spiking activity is part of a neural code, 
it may be reflected in LFP activity.  Figures 6 and 7
present results of spectral analysis of LFP activity.

Figure 6 shows line plots for activity at a single site and across a
population in one monkey.  Panel a) shows the power during the memory
period (solid) compared to the baseline period (dotted) at a single
site for trials to the preferred direction.  Memory period LFP
activity shows a peak similar to the SU activity. Since the LFP is a
continuous time series, a study of activity for other trial conditions
with lower firing rates  is possible without a loss of resolution.
Panel c) shows the elevated gamma band power is absent during the
memory period in the anti-preferred direction (dotted).  Panels b) and
d) show the same information for a population average in one monkey.
The gamma band peak is broader in the population average 
and distinct from temporal structure at lower frequencies below 25Hz.
Inspection of the individual spectra reveals that this broadening 
is due to variability in the peak frequency at different sites in area
LIP.

Figure 7 shows spectrograms for the activity presented in figure 6.
Panel a) shows the spectrogram for activity at a single site during
trials in the preferred direction and panel c) shows the population
average.  The increase in power during the memory period in a
$70 \pm 20$Hz gamma band is sustained throughout the memory period.  Panels b)
and d)  present activity in the anti-preferred  direction for single
site and a population average.  In both cases there is no increase of
activity during the memory period.   Low
frequency ($20 \pm 5$Hz) beta band activity 
shows some organization with respect to the trial in these plots.
This is presented in figure 15.  The analysis of activity from a second monkey
gives similar results.

\subsection*{Coherency between SU and LFP activity during working memory}

If working memory is a network phenomenon temporal structure in SU and
LFP activity may be coherent during the memory period.  We
investigated this possibility by calculating the coherency between  SU
and LFP activity.

Figure 8 shows the coherency between a single cell and the
simultaneously recorded LFP averaged across all trials to the
preferred direction.   A sharp increase in the coherence between SU
activity and the LFP can be seen at $70 \pm 20$Hz that exceeds 99\%
confidence intervals.  This increase is sustained through the memory
period.  When the coherence is significant the  phase of the coherency
is also well-organized in this band and has a value of zero radians.
This is evidence for phase-locking between the SU and the LFP during
working memory  that is sustained throughout the period.  In
particular, the phase of the  coherency indicates that during working
memory the SU fires at the peak of broad-band 50 -- 90Hz LFP
activity.  In a given window, the phase of the coherency is relatively
constant across frequency indicating that SU and LFP gamma band
activity are synchronous with no time lag between them.  It is
important to note that since the power in each process  is explicitly
normalized for, the increase in coherence  during the memory period
is not related to power increases in either process.  Instead it is a
result of an increase in the predictability of one process given the
other indicating  that one can predict when the SU will fire from LFP
activity more accurately during working memory than during simple
fixation.

This result is typical of a population of cells with memory activity
recorded from  area LIP in two monkeys.  Of the 40 single units
recorded at 33 sites in this study, 18 (45\%) showed significant
coherence during the memory period in the $70 \pm 20$Hz frequency band
($p<0.01$).  All single units with significant coherence with the LFP
contained  significant temporal structure during the memory period.  A
comparison of activity between multiple single units and the LFP
recorded at the same site was not possible in this study as there were
only two pairs of simultaneously  recorded single units with
significant memory period activity.

Since the coherency is well-normalized, the coherency from different
recordings can be averaged to determine the population coherency.
Figure 9 shows the population coherency for one monkey.    The
increase in coherence between the LFP and a simultaneously recorded SU
is significant ($p<0.01$) across a population average, even though the
phase at different locations could be random which would reduce the
population average.  Activity from the second monkey shows the same
results.  The significant population coherency during the memory
period is further evidence for locking of SU activity to temporal
structure in the LFP that is activated by a working memory task.
Moreover this locking occurs  with the same preferred phase in all
recordings from area LIP in this study.  

The LFP is measured by low-pass filtering the extracellular potential
recorded on the tetrode.  Since the extracellular potential also
contains spike events there is the possibility that the estimated
coherency is a result of spectral leakage in which low frequencies
present in the spikes  contaminate the LFP estimate.  To control for
this, we suppressed spike energy in the extracellular potential by
smoothly subtracting a mean waveform at each spike event time.  We
then  low-pass filtered the resulting signal as before and repeated
the analysis.   The estimated coherency was still significant.

We also performed a second control for possible contamination of the
LFP estimate by SU spike events.   The mean  potential in a 200ms
window was calculated  conditional on SU spike event times for all
events during the memory period and compared against those during the
baseline.  Figure 10 panel a) shows the spike-triggered average potential
conditional on SU activity during the baseline period and panel b)
shows the average conditional on SU activity  during the memory period.
95\% confidence intervals are shown with dotted horizontal lines.  The
spike-triggered potential during the memory period shows an
oscillatory component that survives the average indicating that the
phase of the LFP at that frequency is coherent.    The baseline
activity shows little coherent  structure at a high frequencies but
does show significant structure immediately before a spike event.  The
oscillations in the  spike-triggered average potential during the
memory period are not significant but they do have the same frequency
and phase as the coherency in figure 8.   Since these error bars are
constructed in the time domain  they are not suitable to detect
band-limited structure in the frequency-domain.  We present a
frequency-domain alternative to the spike-triggered potential in
figure 11.

The increase in the estimated coherency during the memory period may
be misleading because  the low number of SU spike events during the
baseline period makes the estimate of coherency in that period
unreliable.   We performed a second control to see if the coherency
between SU and LFP activity could be observed in a way that was not
sensitive to small numbers of SU spike events during the baseline
period.   We subtracted mean spike waveforms from the raw data  and
bandpass filtered the raw signal at 50 -- 90Hz using projection
filters.  The bandpassed signal is a complex number with amplitude and
phase and we sampled the phase of this signal at the spike event
times.  Since the phase of LFP activity at SU spike event times is
normalized we can compare the coherency of activity across recordings.
We pooled activity from all single units recorded in both monkeys and
compared the distribution of the phase of the LFP at spike event times
during the baseline period with the memory  period.  Unlike the
measures of association presented above, this measure is not 
to small numbers of SU spike events during the baseline period.

Figure 11 shows the normalized histogram of the phase at all spike
events during the memory period (solid) and the baseline period
(dashed).  The distribution of the phases during the memory period is
peaked at zero phase consistent with the coherency estimate.  It is
significantly different from uniform ($p<0.01$ KS test:  N=6192)as well
as significantly different  from the baseline period ($p<0.01$ KS test:
N=6192).  This indicates that the significant memory period coherency
is not due to the larger number of SU spike events during the memory
period compared to the baseline period.  While it is not surprising
that there is a close relation between SU activity and the
locally-recorded LFP, taken together these results indicate that
there are coherent gamma band dynamics in neuronal activity
during working memory that are not present during simple fixation.

\subsection*{Comparing SU and LFP activity}

Comparing the power in SU and LFP activity at  a particular frequency
band over time and across trial condition provides more information
about these processes and how they code for behavior.

Figure 12 compares SU and LFP activity in the $70 \pm 20$Hz band in
the preferred direction during the task.  Panel a) shows line plots of
the activity for the LFP (solid) and SU  (dashed) at a single site.
Panel b) presents the population average  for LFP (solid) and SU
activity (dashed).   Elevations in SU activity at $70 \pm 20$Hz are
mirrored in the LFP throughout the task.  This provides  evidence that
temporal structure in LFP activity  is elevated during the memory
period with similar dynamics to SU activity.

Figure 13 compares SU and LFP activity in the $70 \pm 20$Hz band
across saccade directions at three different epochs of the trial.
Panel a) shows LFP power from a single site perisaccadically (solid),
during the memory period (dashed) and baseline (dot-dashed).  Panel b)
shows SU power in the same frequency band from the same three periods.
This shows that temporal structure in LFP activity is tuned for the
preferred direction during the memory period in a similar way to SU
activity recorded at the same site.

Figure 14 contains 2d plots presenting the spatiotemporal organization
of SU and LFP activity.  Panel a) shows SU activity at a $70 \pm 20$Hz
band from a single cell and panel b) shows LFP activity in the same
frequency band at the same site.   SU and LFP activity show similar
organization during the task.  The increase in LFP power in the $70
\pm 20$Hz band  during the memory period is significant ($p<0.05$) in 27
of 33 ($82\%$) of  sites recorded from.  The tuning of LFP power  to
the preferred direction during the memory period is significant
($p<0.05$) at 28 of 33 ($85\%$) of sites recorded  from.  

In addition to task-related activity in a $70 \pm 20$Hz band there is
evidence that another $20 \pm 5$Hz frequency band  of the LFP contains
task-related activity.  Figure 15 shows LFP power at $20 \pm 5$Hz and
compares it with  $70 \pm 20$Hz  over time.  Panel a) shows the 70Hz
activity (solid) against the 20Hz activity (dashed) for a single
site, averaged across trials  in the preferred direction.  Panel b)
shows the same activity in a population average for one monkey.   As
noted before $70 \pm 20$Hz power is elevated during working memory.
Additional temporal structure in the LFP in a $20 \pm 5$Hz band
increases in power toward the end of the memory period. This is
present in activity at a single site as well as in the population
average and may be related to preparatory aspects of the task.   The
suppression of 20Hz activity peri-saccadically during saccades 
to
all directions is also visible in activity at a single site and a
population average
and may be
related to movement execution.   
These features are also present in LFP activity from
the second monkey.   The suppression or peri-saccadic activity
compared to memory period  is significant ($p<0.05$) at 33 of 33
($100\%$) sites recorded from.  

SU activity is emphasized as a control
signal in current work to develop a neural prosthesis but progress has
been difficult because stable
SU activity is hard to record using chronic cortical implants.
The richness of LFP activity in
multiple frequency bands, and the ease with which it can be acquired
compared to SU activity, suggests that temporal structure we find in the
LFP may be useful in the control of a neural prosthesis.

\section*{DISCUSSION}

Elevations in SU mean firing rates define sustained memory
activity but how this activity is maintained is unknown.   This study
investigates this issue by examining temporal structure in SU and LFP
activity as well as the relation between them during working memory in
area LIP of macaque parietal cortex.

We begin by discussing a methodological issue and then focus on the
three principal findings of this work: i) SU activity contains
temporal structure indicating the existence
of dynamical memory fields; 
ii) SU and LFP activity are coherent
in the gamma frequency band during working memory and not during
simple fixation; 
and iii)  LFP activity contains temporal structure
that codes for movement planning and
execution and may be useful for the control of a neural prosthesis.

\subsection*{General methodological issue:  Time and frequency}

The presence of temporal structure in neural activity is of much
interest \cite{SingerGray1995} and previous work in our lab has
investigated it in these data using correlation function  measures
\cite{PezarisSahaniAndersen1997b,PezarisSahaniAndersen1999}.
Correlation functions are often estimated in the time domain to detect
temporal structure but are well known to suffer serious problems of
bias and variance which are exacerbated in the context of a behaving
animal \cite{JarvisMitra2000}.  Attempts to  overcome these
limitations when estimating correlation functions 
by pooling observations across a large period of time
during the experiment 
leads to a
violation of the stationarity assumption and misinterpretation of the
data \cite{Brody1999b}.

Here we resolve significant temporal structure in neuronal activity in
parietal cortex using spectral analysis.  
Spectral quantities are estimated in the frequency domain and while
they are mathematically equivalent to correlation  functions their
statistical properties are better understood and better estimated.
The problems of estimation bias and variance are controlled  by using
modern methods of multitaper spectral  analysis \cite{Thomson1982,PercivalWalden1993}.  Other  advantages of working in the
frequency domain are that i) weak non-stationarity only manifests
itself in the spectrum at low frequencies; ii) nearby points in
frequency are statistically independent resulting in local error  bars
for the estimates; and iii) the problem of the normalization of the
cross-correlation function is addressed by using the coherency which
is dimensionless.  

In recent work, Mitra and Pesaran (1999) 
has applied multitaper spectral estimation to
a wide range of neural data and  demonstrates their advantages.   In
general, these are most pronounced when dealing with short segments of
possibly nonstationary data.  Such problems are particularly  severe
when studying neural activity in a behaving animal performing  a
structured task.  Consequently,  we believe the inappropriate  use of
time-domain correlation function measures may explain the failure of
previous attempts to detect temporal structure in these data.
However, if one did want to evaluate time-domain correlations, the
optimal way to estimate them would be to inverse Fourier transform the
corresponding spectral quantities \cite{PezarisSahaniAndersen2000}.

\subsection*{Understanding SU activity:  Dynamical memory
fields of spectral structure}

An important aspect of the finding is that the spectrum of SU activity
demonstrates the presence of temporal structure during working
memory that comprise dynamical memory fields.  
Mathematically, SU  activity is a point process composed of
discrete events in time (action potentials) in contrast to continuous
processes such as the LFP that consist of continuous voltage changes.
The simplest  point process is the Poisson process which is often used
to model sequences of action potentials that seem to occur randomly in
time.  This process is parameterized by a rate: the average number of
events occurring during a given interval \cite{CoxLewis1966}.   It has
the important property that the events are statistically independent.
This means that  the probability of an event occuring in a given
interval is not dependent on activity  before or after the interval.
The estimation of the rate parameter also reflects this because the
number of events during an interval does not depend on their  ordering
in time.  Thus, there is no temporal structure in the sequence of
events in a Poisson process.

In contrast to the mean firing rate, the spectrum depends on the
ordering of events in time and the process is not assumed to be
random.  The key difference between the spectrum of a continuous
process and that of a point process is that a point process spectrum
has a  non-zero high frequency limit \cite{Bartlett1963}.  If events
in a process  occur randomly, as for a Poisson process, the spectrum
of the process is flat and uniform indicating a lack of temporal
structure.  A SU spectrum  with significant deviations from uniformity
indicates that  the underlying
process is not Poisson with constant rate and
is therefore  evidence for temporal structure in SU activity.
Importantly, the mean
firing rate and the spectrum for the same interval do not capture the
same information.  In fact, in general 
the estimate of the mean firing rate does
not predict the frequency at the  peak of the spectrum.  The
only property of the spectrum that the mean firing rate does
characterize is the high frequency limit, or equivalently for a
Poisson process, the magnitude.    Our results indicate that there is
broad-band temporal structure in SU activity that defines a location
in space and is organized in dynamical memory fields.

The significant spectral structure we observe in SU activity has
interesting consequences for classes of the potential underlying
processes.  For example, the suppression at low frequencies reflects
an effective refractory period for the cell of approximately 10ms.
This is not related to the biophysical refractory period and
represents a ``2nd refractory period'' during memory that may reflect
a characteristic integration time for the cell.   The presence of
spectral suppression  means that even a Poisson process  with a
time-varying rate cannot describe the dynamics of SU memory activity
\cite{Brillinger1978} and other classes of models for point process
activity must be considered.

\subsection*{Coherent gamma band network activity during working
memory}

We find that during the memory period the coherency between SU and LFP
activity is significant in the gamma band, 50 -- 90Hz.
Moreover, SUs exhibit  phase locking to the LFP and preferentially
fire at the peak of an LFP oscillation throughout the memory period.
This temporal structure is not present during simple fixation.  The
result is confirmed by two controls.  The first is a spike-triggered
average of the extracellular potential that controls for spurious
coherency due to spectral leakage of SU  activity into the LFP.  The
second is a histogram of the phase of the LFP in the gamma band at
spike times that controls for the sensitivity of the coherency
estimate to small numbers of spikes during the baseline period.

The coherent structure we observe is similar to neural activity
underlying the temporal code present during odor presentation in the locust
\cite{WehrLaurent1996} and provides evidence for dynamical memory
fields of spectral structure that code for movement plans.  
The relation of SU activity to the LFP has
been previously characterized in V1 of the monkey
\cite{Livingstone1996} and in sensorimotor cortex of macaques during
exploratory and trained motor behavior \cite{MurthyFetz1996a,MurthyFetz1996b,DonoghueSanesHatsopoulos1998}.  In
particular, the latter studies found relations between  the processes
during preparatory aspects of the task.  This is possibly related to
the activity we observe in the LFP in the 15 -- 25Hz range, since our
observations of the dynamics of
spectral power in that band are consistent with a preparatory signal.

Gamma band activity has been previously reported in the LFP of cats
performing a sensorimotor task \cite{RougeulBouyerDedet1979} and
sustained gamma band activity has been reported in human EEG studies
during a delayed-response task
\cite{Tallon-baudryBertrandPeronnet1998,Tallon-baudryKreiterBertrand1999}.  These and other reports have
led to the suggestion that gamma band activity may be important for
cognitive processing and consciousness
\cite{LlinasRibaryPedroarena1999}.  Investigating gamma
band coherency between SUs and the LFP during working memory in a
macaque may help
bridge the gap between results from EEG and SU activity and offers an
opportunity to study the neural substrate underlying potentially
related  cognitive processes in human and non-human primates
\cite{Tallon-baudryBertrand1999}.  In particular,
dynamical memory fields may reflect the columnar organization of
cortex and could provide insight into the role of long-range
reverberating circuits in the formation and sustainance of working
memory activity.

\subsection*{LFP activity codes for movement planning and execution}

The parietal cortex is implicated in movement planning, and SU
activity in area LIP codes for intended eye movements and arm
movements in a retinotopic coordinate frame \cite{Andersen1995,BracewellMazzoniAndersen1996,SnyderBatistaAndersen1997}.
Recently, it has been proposed that goal-oriented retinotopic activity
in parietal cortex  may be useful for the control of a prosthesis
\cite{ShenoyKureshiAndersen1999}.  However, the difficulty of 
isolating activity from SUs with cortical implants could be a
significant hurdle in the development of this application.

We show gamma band activity in the LFP mirrors SU activity during
movement planning.  Two other aspects of LFP activity are also related
to the task.  Firstly, the tuning of both LFP mean response and
activity in the gamma band broadens peri- and post-saccadically.  The
significance of this is not known, but it may be related to signals to
update the eye fields in area LIP following the  saccade
\cite{DuhamelColbyGoldberg1992,BatistaBuneoAndersen1999}.
Secondly, the LFP has additional temporal structure in a distinct
15 -- 25Hz frequency band with complicated dynamics that could be related
to movement execution  and preparatory aspects of the task.  These
results indicate the LFP contains a rich variety of information about the
activation of parietal cortex during movement planning and execution.

Our results are evidence for coherent neuronal dynamics
in SU and LFP activity during working memory organized in dynamical memory 
fields.    The presence of
significant 
task-related spectral structure provides evidence for a temporal structure
in SU and LFP activity in parietal cortex.  
Since the LFP is simpler to acquire than SU
activity, the presence of temporal structure in LFP activity may be
useful for the control of a neural prosthesis.


\section*{Appendix}

Here we present modern multitaper  methods of spectral analysis used
in this paper.   These methods were introduced in \cite{Thomson1982}
and have been succesfully applied to  neurobiological data in recent
work \cite{MitraPesaran1999,PrechtlCohenKleinfeld1997,CacciatoreBrodfuehrerKleinfeld1999,MitraOgawaUgurbil1997}.  
Multitaper methods involve
the use of multiple data tapers for spectral estimation.  A variety of
tapers can be used, but an optimal family of orthogonal tapers is
given by the prolate spheroidal functions or Slepian functions.  
These are parameterized by
their length in time, $T$, and their bandwidth in frequency, $W$
\cite{SlepianPollack1961}.  For choice of $T$ and $W$, up to $K=2TW-1$
tapers are concentrated in frequency and suitable for use in spectral
estimation.

The ordinary continuous-valued time series and point processes are
considered in this work form a hybrid data set and spectral analysis
provides a unified framework for their analysis.
For the ordinary time series consider a 
continuous-valued process, $x_t, t = 1, \ldots, N$.  
The basic quantity
for further analysis is the windowed Fourier transform, $\tilde x_k(f)$:

\begin{equation}
\tilde x_k(f) = \sum_1^N w_t(k) x_t exp (-2\pi i f t) 
\end{equation}

\noindent $w_t(k)~ (k=1,2,\ldots,K)$ are $K$  orthogonal taper functions.  

For the point process consider a sequence of event times
$\{\tau_j\},
j=1, \ldots, N $ in the interval $[0,T]$.  The quantity for further 
analysis of point processes is also
the windowed Fourier transform, denoted by $\tilde x_k(f)$:

\begin{equation}
\tilde x_k(f) = \sum_{j=1}^N w_{\tau_j}(k) exp (-2\pi i f \tau_j) 
-{N(T) \over T} \tilde w_0(k) 
\end{equation}

\noindent $w_0(k)$ is the Fourier transform of the data taper at zero
frequency and $N(T)$ is the total number of spikes in the interval.

When averaging over trials we introduce an additional
index, $i$, denoting trial number, $\tilde x_{k,i}(f)$.

When dealing with either point or continuous processes, 
the multitaper estimates for the spectrum $S_{x}(f)$, cross-spectrum
$S_{yx}(f)$ and coherency $C_{yx}(f)$ are given by

\begin{equation}
S_{x}(f) = {1\over K} \sum_{k=1}^K |\tilde x_k(f)|^2 
\end{equation}

\begin{equation}
S_{yx}(f) =  {1\over K} \sum_{k=1}^K \tilde  y_k(f) \tilde  x_k^*(f) 
\end{equation}

\begin{equation}
C_{yx}(f) =  {  S_{yx}(f)
\over
\sqrt{ S_{x}(f) S_{y}(f)}} 
\label{mtaper_coh}
\end{equation}

The auto- and cross-correlation functions  can be obtained by inverse 
Fourier transforming the spectrum and cross-spectrum.

\newpage
\bibliography{MemSacLIP}

\begin{thebibliography}{}

\bibitem[Amit, 1995]{Amit1995}
Amit, D. (1995).
\newblock The hebbian paradigm reintegrated: local reverberations as internal
  representation.
\newblock {\em Behav. Brain Sci.}, 18:617--626.

\bibitem[Andersen, 1995]{Andersen1995}
Andersen, R. (1995).
\newblock Encoding of intention and spatial location in the posterior parietal
  cortex.
\newblock {\em Cerebral Cortex}, 5(5):457--469.

\bibitem[Baddeley, 1992]{Baddeley1992}
Baddeley, A. (1992).
\newblock Working memory.
\newblock {\em Science}, 255:556--559.

\bibitem[Barash et~al., 1991]{BarashBracewellAndersen1991a}
Barash, S., Bracewell, R., Fogassi, L., Gnadt, J., and Andersen, R. (1991).
\newblock Saccade-related activity in the lateral intraparietal area .1.
  temporal properties - comparison with area 7a.
\newblock {\em J. Neurophysiol.}, 66:1095--1108.

\bibitem[Bartlett, 1963]{Bartlett1963}
Bartlett, M. (1963).
\newblock The spectral analysis of point processes.
\newblock {\em J. R. Stat. Soc. Ser. B (Meth)}, 25:264--269.

\bibitem[Batista et~al., 1999]{BatistaBuneoAndersen1999}
Batista, A., Buneo, C., Snyder, L., and Andersen, R. (1999).
\newblock Reach plans in eye-centered coordinates.
\newblock {\em Science}, 285(5425):257--260.

\bibitem[Borst and Theunissen, 1999]{BorstTheunissen1999}
Borst, A. and Theunissen, F. (1999).
\newblock Information theory and neural coding.
\newblock {\em Nature Neurosci.}, 2(11):947--957.

\bibitem[Bracewell et~al., 1996]{BracewellMazzoniAndersen1996}
Bracewell, R., Mazzonni, P., Barash, S., and Andersen, R. (1996).
\newblock Motor intention activity in the macaque's lateral intraparietal area
  .2. changes of motor plan.
\newblock {\em J. Neurophysiol.}, 76:1457--1464.

\bibitem[Bressler et~al., 1993]{BresslerCoppolaNakamura1993}
Bressler, S., Coppola, R., and Nakamura, R. (1993).
\newblock Episodic multiregional cortical coherence at multiple frequencies
  during visual task-performance.
\newblock {\em Nature}, 366:153--156.

\bibitem[Brillinger, 1975]{Brillinger1975}
Brillinger, D. (1975).
\newblock {\em Time series data analysis and theory}.
\newblock Holt, Rinehart and Winston, Inc:New York.

\bibitem[Brillinger, 1978]{Brillinger1978}
Brillinger, D. (1978).
\newblock {\em Developments in Statistics}, volume~1, pages 33--129.
\newblock Academic Press Inc.

\bibitem[Brody, 1999]{Brody1999b}
Brody, C. (1999).
\newblock Correlations without synchrony.
\newblock {\em Neural Comput.}, 11(7):1537--1551.

\bibitem[Bruce and Goldberg, 1985]{BruceGoldberg1985}
Bruce, C. and Goldberg, M. (1985).
\newblock Primate frontal eye fields. i. single neurons discharging before
  saccades.
\newblock {\em J. Neurophysiol.}, 53:603--635.

\bibitem[Cacciatore et~al., 1999]{CacciatoreBrodfuehrerKleinfeld1999}
Cacciatore, T., Brodfuehrer, P., Gonzalez, J., Jiang, T., Adams, S., Tsien, R.,
  Kristan, W., and Kleinfeld, D. (1999).
\newblock Identification of neural circuits by imaging coherent electrical
  activity with fret-based dyes.
\newblock {\em Neuron}, 23(3):449--459.

\bibitem[Chafee and Goldman-Rakic, 1998]{ChafeeGoldman-Rakic1998}
Chafee, M. and Goldman-Rakic, P. (1998).
\newblock Matching patterns of activity in primate prefrontal area 8a and
  parietal area 7ip neurons during a spatial working memory task.
\newblock {\em J. Neurophysiol.}, 79(6):2919--2940.

\bibitem[Chapin et~al., 1999]{ChapinMoxonNicolelis1999}
Chapin, J., Moxon, K., Markowitz, R., and Nicolelis, M. (1999).
\newblock Real-time control of a robot arm using simultaneously recorded
  neurons in the motor cortex.
\newblock {\em Nature Neurosci.}, 2(7):664--670.

\bibitem[Cox and Lewis, 1966]{CoxLewis1966}
Cox, D. and Lewis, P. (1966).
\newblock {\em The statistical analysis of series of events}.
\newblock Chapman and Hall:London.

\bibitem[deCharms and Merzenich, 1996]{deCharmsMerzenich1996}
deCharms, R. and Merzenich, M. (1996).
\newblock Primary cortical representation of sounds by the coordination of
  action-potential timing.
\newblock {\em Nature}, 381:610--613.

\bibitem[Donoghue et~al., 1998]{DonoghueSanesHatsopoulos1998}
Donoghue, J., Sanes, J., Hatsopoulos, N., and Gaal, G. (1998).
\newblock Neural discharge and local field potential oscillations in primate
  motor cortex during voluntary movements.
\newblock {\em J. Neurophysiol.}, 79(1):159--173.

\bibitem[Duhamel et~al., 1992]{DuhamelColbyGoldberg1992}
Duhamel, J., Colby, C., and Goldberg, M. (1992).
\newblock The updating of the representation of visual space in parietal cortex
  by intended eye-movements.
\newblock {\em Science}, 255(5040):90--92.

\bibitem[Eckhorn et~al., 1988]{EckhornBauerReitboeck1988}
Eckhorn, R., Bauer, R., Jordan, W., Brosch, M., Kruse, W., Munk, M., and
  Reitboeck, H. (1988).
\newblock Coherent oscillations - a mechanism of feature linking in the visual
  cortex - multiple electrode and correlation analyses in the cat.
\newblock {\em Biol. Cybern.}, 60(2):121--130.

\bibitem[Engel et~al., 1990]{EngelKonigSinger1990}
Engel, A., Konig, P., Gray, C., and Singer, W. (1990).
\newblock Stimulus-dependent neuronal oscillations in cat visual-cortex -
  intercolumnar interaction as determined by cross-correlation analysis.
\newblock {\em Eur. J. Neurosci.}, 2:588--606.

\bibitem[Funahashi et~al., 1989]{FunahashiBruceGoldman-Rakic1989}
Funahashi, S., Bruce, C., and Goldman-Rakic, P. (1989).
\newblock Mnemonic coding of visual space in the monkey's dorsolateral
  prefrontal cortex.
\newblock {\em J. Neurophysiol.}, 61:331--349.

\bibitem[Fuster, 1995]{Fuster1995}
Fuster, J. (1995).
\newblock {\em Memory in cerebral cortex: An empirical approach to neural
  networks in the human and nonhuman brain}.
\newblock MIT Press:Cambridge, MA.

\bibitem[Fuster and Jervey, 1982]{FusterJervey1982}
Fuster, J. and Jervey, J. (1982).
\newblock Neuronal firing in the inferotemporal cortex of the monkey in a
  visual memory task.
\newblock {\em J. Neurosci.}, 2(3):361--375.

\bibitem[Gnadt and Andersen, 1988]{GnadtAndersen1988}
Gnadt, J. and Andersen, R. (1988).
\newblock Memory related motor planning activity in posterior parietal cortex
  of macaque.
\newblock {\em Exp. Brain Res.}, 70:216--220.

\bibitem[Goldman-Rakic, 1988]{Goldman-Rakic1988}
Goldman-Rakic, P. (1988).
\newblock Topography of cognition - parallel distributed networks in primate
  association cortex.
\newblock {\em Ann. Rev. Neurosci.}, 11:137--156.

\bibitem[Goldman-Rakic, 1995]{Goldman-Rakic1995}
Goldman-Rakic, P. (1995).
\newblock Cellular basic of working memory.
\newblock {\em Neuron}, 14(3):477--485.

\bibitem[Gray et~al., 1989]{GrayKonigSinger1989}
Gray, C., Konig, P., Engel, A., and Singer, W. (1989).
\newblock Oscillatory responses in cat visual-cortex exhibit inter-columnar
  synchronization which reflects global stimulus properties.
\newblock {\em Nature}, 338:334--337.

\bibitem[Hebb, 1949]{Hebb1949}
Hebb, D. (1949).
\newblock {\em Organization of Behavior}.
\newblock Wiley:New York.

\bibitem[Jarvis and Mitra, 2000]{JarvisMitra2000}
Jarvis, M. and Mitra, P. (2000).
\newblock Sampling properties of the spectrum and coherency of sequences action
  potentials.
\newblock {\em Neural Comput.}, Submitted.

\bibitem[Kennedy and Bakay, 1998]{KennedyBakay1998}
Kennedy, P. and Bakay, R. (1998).
\newblock Restoration of neural output from a paralyzed patient by direct brain
  connection.
\newblock {\em Neuroreport}, 9:1707--1711.

\bibitem[Koch and Fuster, 1989]{KochFuster1989}
Koch, K. and Fuster, J. (1989).
\newblock Unit-activity in monkey parietal cortex related to haptic perception
  and temporary memory.
\newblock {\em Exp. Brain Res.}, 76(2):292--306.

\bibitem[Livingstone, 1996]{Livingstone1996}
Livingstone, M. (1996).
\newblock Oscillatory firing and interneuronal correlations in squirrel monkey
  striate cortex.
\newblock {\em J. Neurophysiol.}, 75(6):2467--2485.

\bibitem[Llinas et~al., 1999]{LlinasRibaryPedroarena1999}
Llinas, R., Ribary, U., Contreras, D., and Pedroarena, C. (1999).
\newblock The neuronal basis for consciousness.
\newblock {\em Phil. Trans. R. Soc. Lond. Ser. B - Biol. Sci.}, 353:1841--1849.

\bibitem[Lorente~de No, 1938]{LorentedeNo1938}
Lorente~de No, R. (1938).
\newblock {\em Physiology of the nervous system}, chapter Cerebral cortex
  architecture, intracortical connections, motor projections, pages 291--339.
\newblock Oxford University Press, Oxford.

\bibitem[Miller et~al., 1996]{MillerEricksonDesimone1996}
Miller, E., Erickson, C., and Desimone, R. (1996).
\newblock Neural mechanisms of visual working memory in prefrontal cortex of
  the macaque.
\newblock {\em J. Neurosci.}, 16:5154--5167.

\bibitem[Mitra et~al., 1997]{MitraOgawaUgurbil1997}
Mitra, P., Ogawa, S., Hu, X., and Ugurbil, K. (1997).
\newblock The nature of spatiotemporal changes in cerebral hemodynamics as
  manifested in functional magnetic resonance imaging.
\newblock {\em Mag. Res. Med.}, 37:511--518.

\bibitem[Mitra and Pesaran, 1999]{MitraPesaran1999}
Mitra, P. and Pesaran, B. (1999).
\newblock Analysis of dynamic brain imaging data.
\newblock {\em Biophys. J.}, 76:691--708.

\bibitem[Miyashita and Chang, 1988]{MiyashitaChang1988}
Miyashita, Y. and Chang, H. (1988).
\newblock Neuronal correlate of pictorial short-term memory in the primate
  temporal cortex.
\newblock {\em Nature}, 331:68--70.

\bibitem[Murthy and Fetz, 1996a]{MurthyFetz1996a}
Murthy, V. and Fetz, E. (1996a).
\newblock Oscillatory activity in sensorimotor cortex of awake monkeys:
  Synchronization of local field potentials and relation to behavior.
\newblock {\em J. Neurophysiol.}, 76(6):3949--3967.

\bibitem[Murthy and Fetz, 1996b]{MurthyFetz1996b}
Murthy, V. and Fetz, E. (1996b).
\newblock Synchronization of neurons during local field potential oscillations
  in sensorimotor cortex of awake monkeys.
\newblock {\em J. Neurophysiol.}, 76(6):3968--3982.

\bibitem[Percival and Walden, 1993]{PercivalWalden1993}
Percival, D. and Walden, A. (1993).
\newblock {\em Spectral analysis for physical applications}.
\newblock Cambridge University Press, Cambridge, UK.

\bibitem[Pezaris, 2000]{Pezaris2000}
Pezaris, J. (2000).
\newblock {\em Local circuitry in LIP}.
\newblock PhD thesis, California Institute of Technology.

\bibitem[Pezaris et~al., 1997a]{PezarisSahaniAndersen1997b}
Pezaris, J., Sahani, M., and Andersen, R. (1997a).
\newblock Extracellular recording from adjacent neurons: Ii. correlations in
  macaque parietal cortex.
\newblock In {\em Soc. Neurosci. Abs.}, volume~23.

\bibitem[Pezaris et~al., 1997b]{PezarisSahaniAndersen1997a}
Pezaris, J., Sahani, M., and Andersen, R. (1997b).
\newblock Tetrodes for monkeys.
\newblock In Bower, J., editor, {\em Computational Neuroscience}. Plenum
  Press:New York.

\bibitem[Pezaris et~al., 1999]{PezarisSahaniAndersen1999}
Pezaris, J., Sahani, M., and Andersen, R. (1999).
\newblock Response-locked changes in auto- and cross-covariations in parietal
  cortex.
\newblock {\em Neurocomp.}, 26-27:471--476.

\bibitem[Pezaris et~al., 2000]{PezarisSahaniAndersen2000}
Pezaris, J., Sahani, M., and Andersen, R. (2000).
\newblock Spike train coherence in macaque parietal cortex during a memory
  saccade task.
\newblock {\em Neurocomp.}, In press.

\bibitem[Prechtl et~al., 1997]{PrechtlCohenKleinfeld1997}
Prechtl, J., Cohen, L., Pesaran, B., Mitra, P., and Kleinfeld, D. (1997).
\newblock Visual stimuli induce waves of electrical activity in turtle cortex.
\newblock {\em Proc. Natl. Acad. Sci. USA}, 94(14):7621--7626.

\bibitem[Rao, 1965]{Rao1965}
Rao, C. (1965).
\newblock {\em Linear Statistical Inference and its Applications}.
\newblock Wiley:New York.

\bibitem[Reece and O'Keefe, 1989]{ReeceO'Keefe1989}
Reece, M. and O'Keefe, J. (1989).
\newblock The tetrode: An improved technique for multi-unit extracellular
  recording.
\newblock In {\em Soc. Neurosci. Abs.}, volume~15.

\bibitem[Roelfsema et~al., 1997]{RoelfsemaEngelSinger1997}
Roelfsema, P., Engel, A., Konig, P., and Singer, W. (1997).
\newblock Visuomotor integration is associated with zero time-lag
  synchronization among cortical areas.
\newblock {\em Nature}, 385(6612):157--161.

\bibitem[Rosenberg et~al., 1989]{RosenbergAmjadHalliday1989}
Rosenberg, J., Amjad, A., Breeze, P., Brillinger, D., and Halliday, D. (1989).
\newblock The fourier approach to the indentification of functional coupling
  between neuronal spike trains.
\newblock {\em Prog. Biophys. Molec. Biol.}, 53:1--31.

\bibitem[Rouguel et~al., 1979]{RougeulBouyerDedet1979}
Rouguel, A., Bouyer, J., Dedet, L., and Debray, O. (1979).
\newblock Fast somato-parietal rhythms during combined focal attention and
  immobility in baboon and squirrel monkey.
\newblock {\em Electro. Clin. Neurophysiol.}, 46:310--319.

\bibitem[Sahani, 1999]{Sahani1999}
Sahani, M. (1999).
\newblock {\em Latent variable models for neural data analysis}.
\newblock PhD thesis, California Institute of Technology.

\bibitem[Sahani et~al., 1998]{SahaniPezarisAndersen1998}
Sahani, M., Pezaris, J., and Andersen, R. (1998).
\newblock On the separation of signals from neighboring cells in tetrode
  recordings.
\newblock In Jordan, M., Kearns, M., and Solla, S., editors, {\em Advances in
  Neural Information Processing Systems 10}. MIT Press:Cambridge, MA.

\bibitem[Sakata et~al., 1995]{SakataTairaMurata1995}
Sakata, H., Taira, M., Murata, A., and Mine, S. (1995).
\newblock Neural mechanisms for the visual guidance of hand action in the
  parietal cortex of the monkey.
\newblock {\em Cereb. Cortex.}, 5:429--438.

\bibitem[Sanes and Donoghue, 1993]{SanesDonoghue1993}
Sanes, J. and Donoghue, J. (1993).
\newblock Oscillations in local-field potentials of the primate motor cortex
  during voluntary movement.
\newblock {\em Proc. Natl. Acad. Sci. USA}, 90:4470--4474.

\bibitem[Seung, 1996]{Seung1996}
Seung, H. (1996).
\newblock How the brain keeps the eyes still.
\newblock {\em Proc. Natl. Acad. Sci. USA}, 93:13339--13344.

\bibitem[Shenoy et~al., 1999]{ShenoyKureshiAndersen1999}
Shenoy, K., Kureshi, S., Meeker, D., Gillikin, B., Batista, A., Buneo, C., Cao,
  S., Burdick, J., and Andersen, R. (1999).
\newblock Toward prosthetic systems controlled by parietal cortex.
\newblock In {\em Soc. Neurosci. Abs.}, volume~25.

\bibitem[Singer and Gray, 1995]{SingerGray1995}
Singer, W. and Gray, C. (1995).
\newblock Visual feature integration and the temporal correlation hypothesis.
\newblock {\em Ann. Rev. Neurosci.}, 18:555--586.

\bibitem[Slepian and Pollack, 1961]{SlepianPollack1961}
Slepian, D. and Pollack, H. (1961).
\newblock Prolate spheroidal wavefunctions. fourier analysis and uncertainty i.
\newblock {\em Bell Sys. Tech. J.}, 40:43--63.

\bibitem[Snyder et~al., 1997]{SnyderBatistaAndersen1997}
Snyder, L., Batista, A., and Andersen, R. (1997).
\newblock Coding of intention in the posterior parietal cortex.
\newblock {\em Nature}, 386(6621):167--170.

\bibitem[Tallon-Baudry and Bertrand, 1999]{Tallon-baudryBertrand1999}
Tallon-Baudry, C. and Bertrand, O. (1999).
\newblock Oscillatory gamma activity in humans and its role in object
  representation.
\newblock {\em Trends Cog. Sci.}, 3(4):151--162.

\bibitem[Tallon-Baudry et~al., 1998]{Tallon-baudryBertrandPeronnet1998}
Tallon-Baudry, C., Bertrand, O., Peronnet, F., and Pernier, J. (1998).
\newblock Induced gamma-band activity during the delay of a visual short-term
  memory task in humans.
\newblock {\em J. Neurosci.}, 18(11):4244--4254.

\bibitem[Tallon-Baudry et~al., 1999]{Tallon-baudryKreiterBertrand1999}
Tallon-Baudry, C., Kreiter, A., and Bertrand, O. (1999).
\newblock Sustained and transient oscillatory responses in the gamma and beta
  bands in a visual short-term memory task in humans.
\newblock {\em Vis. Neurosci.}, 16(3):449--459.

\bibitem[Thomson, 1982]{Thomson1982}
Thomson, D. (1982).
\newblock Spectrum estimation and harmonic analysis.
\newblock {\em Proc. IEEE}, 70:1055--1096.

\bibitem[Thomson, 1994]{Thomson1994c}
Thomson, D. (1994).
\newblock Projection filters for data analysis.
\newblock In {\em Proc. Seventh IEEE SP Work. Stat. Sig. and Array Proc.},
  pages 39--42, Quebec, Canada.

\bibitem[Thomson and Chave, 1991]{ThomsonChave1991}
Thomson, D.~J. and Chave, A.~D. (1991).
\newblock {\em Advances in Spectrum Analysis and Array Processing}, volume~1,
  pages 58--113.
\newblock Prentice Hall.

\bibitem[Ueda and Nakano, 1994]{UedaNakano1994}
Ueda, N. and Nakano, R. (1994).
\newblock Mixture density estimation via em algorithm with deterministic
  annealing.
\newblock {\em Proc. IEEE: Neural networks and signal processing}, 69:69--77.

\bibitem[Wang, 1999]{Wang1999}
Wang, X.-J. (1999).
\newblock Synaptic basis of cortical persistent activity: the importance of
  nmda receptors to working memory.
\newblock {\em J. Neurosci.}, 19:9587--9603.

\bibitem[Wehr and Laurent, 1996]{WehrLaurent1996}
Wehr, M. and Laurent, G. (1996).
\newblock Odour encoding by temporal sequences of firing in oscillating neural
  assemblies.
\newblock {\em Nature}, 384:162--166.

\bibitem[Zhou and Fuster, 1996]{ZhouFuster1996}
Zhou, Y. and Fuster, J. (1996).
\newblock Mnemonic neuronal activity in somatosensory cortex.
\newblock {\em Proc. Natl. Acad. Sci. USA}, 93(19):10533--10537.

\end{thebibliography}
\bibliographystyle{apalike}

\newpage
\thefignotes

\efig{ {\bf The memory-saccade task }  The monkey performs a memory-saccade
to one of eight saccade directions.  a) The trial begins with the illumination
of a fixation light at the center of the screen.  The monkey saccades to the
fixation light. to which the monkey maintains
fixation for one second, which determines the baseline period.  b) A
target is then flashed in one of eight points for 100ms and extinguished.
c) The monkey must maintain fixation for a further second at which point
d) the fixation light is extinguished and the monkey performs a saccade
to the remembered target location.  When the saccade is completed and the
monkey's eye position is within $2\dgr$ of the target, e) it reilluminates
and the monkey is rewarded with a drop of juice.  A corrective minisaccade
follows the target reillumination.  There is a short intertrial interval
before the fixation light reilluminates signalling the start of a new trial.}{ 
  \centerline{\ps{3.4in}{!}{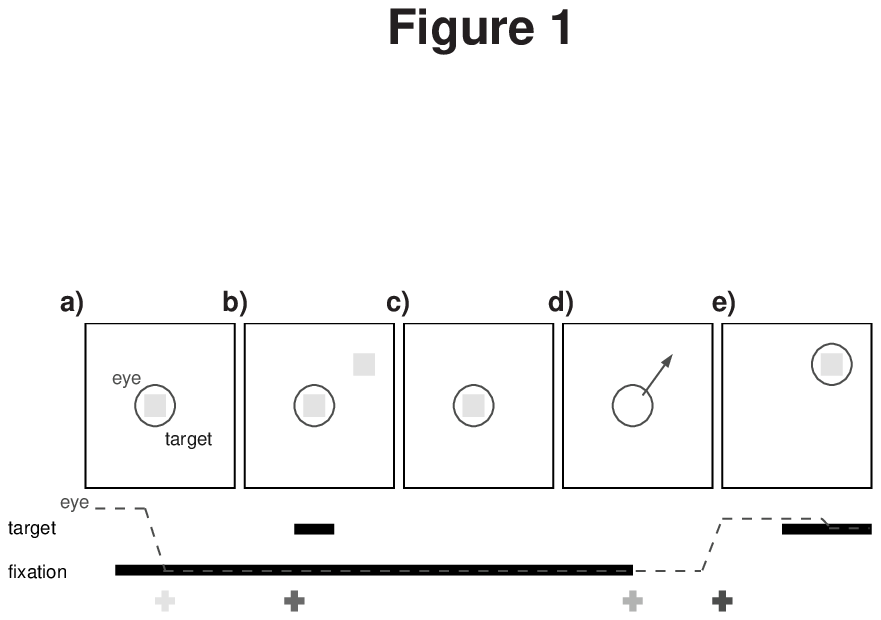}}
  }{fig:TheTask}

\efig{ 
{\bf Sample data }  a)  One channel of raw tetrode data sampled 
at 20kHz shown for a memory-saccade trial during a saccade to the preferred
direction.  The saccade was down and to the left and (x,y) eye position 
traces are shown beneath.  Behavioral events are indicated below eye position
traces establishing the convention to be used in later figures.  Dotted lines
indicated the illumination of the fixation light.  Solid lines indicate
illumination of the target.  The points indicate saccades.  When
available, color is used in place of symbols.  Blue indicates illumination
of the fixation light, red indicates illumination of the target
and green marks the saccade.
These
raw traces were high-pass filtered and threshholded to determine spike event
times.  These events were then classified to give point processes of single 
unit activity.   Note increased activity during the memory period.
b)  The data in panel a) viewed on an expanded time base from 0.2--0.5s.  
Broad-band low frequency ($<100$Hz) 
and SU spiking activity are visible.  
The amplitude of spikes from SU activity is not large compared
to the amplitude of the broad-band activity.
}{
\centerline{\ps{3.4in}{!}{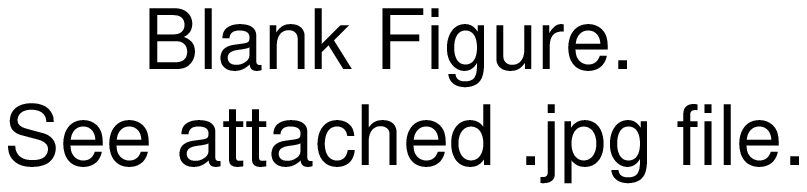}}
}{
fig:RawData
}

\efig{ 
{\bf SU activity } The PSTH is shown for a single unit
recorded during the memory saccade task as a function of 
saccade direction.  A strong increase in single cell activity
during the memory period is seen for saccades to the left.
This defines the preferred direction of the cell.  Eye position
traces, x and y coordinates, for each trial are shown beneath each panel.
Behavioral events during the trial are shown beneath
each histogram according to the convention described for Figure 2.
Spike rasters are shown for each trial at the top of each panel.  
}{ 
\centerline{\ps{3.4in}{!}{Blank.eps}}
}{
fig:SingleUnitActivity
}

\efig{  
{\bf SU and LFP mean response}
Line and 2d plots showing the mean response of SU and LFP activity 
averaged across saccades in each direction.

a) 2d plot of the SU PSTH (as seen in figure 3)  averaged
across trials aligned to the initial target onset 
as a function of saccade direction.  Time is on the x-axis
and saccade direction, aligned with 0 degrees the preferred direction, is
on the y-axis.  
b) Line plots of SU mean response in panel a) 
in preferred (solid) and anti-preferred (dashed) directions.
c) 2d plot of the LFP mean response averaged
across trials aligned to the initial target onset 
as a function of saccade direction.
Time is on the x-axis
and saccade direction, aligned with 0 degrees the preferred direction, is
on the y-axis.  
d)  Line plot of the LFP mean response in the preferred (solid) and
anti-preferred (dashed) directions.

SU and LFP mean responses contain tuned activity.
The LFP mean response also contains tuned 
low frequency postsaccadic activity following saccades made to all
directions.  
}{ 
\centerline{\ps{3.4in}{!}{Blank.eps}}
}{
fig:SU.EvokedLFP
}

\efig{ {\bf Spectral structure in SU activity}
Line and 2d plots showing the spectrum of SU activity in the preferred
direction.

a) Line plots of the spectrum during the
memory (red) and baseline (blue) period for a single cell.
Solid lines indicate the trial-average spectrum.  Dashed lines indicate
95\% error bars estimated using the jacknife.
Dotted lines indicate the high-frequency limit.  
b) 2d plot of the spectrogram for a single cell with time on the
x-axis and frequency on the y-axis.  Power is color-coded
on a log scale. Behavioral events are indicated below.
c) Line plots of the population average spectrum during the
memory (red) and baseline (blue) period for one monkey.
Solid lines indicate the spectrum and dotted lines indicate the 
high-frequency limit.  
d) 2d plot of the population average spectrogram with time on the
x-axis and frequency on the y-axis.  Power is color-coded
on a log scale.  Behavioral events are indicated below.

Significant spectral structure is present during the memory period and
not during the baseline period.  During the memory period a 
significant peak in the spectrum is present 
above 30Hz and significant suppression of the spectrum 
is present below 20Hz for a single cell and in the 
population averages.  This means that SU activity cannot be well
modelled as a rate-varying, inhomogeneous Poisson process.
Spectrograms show that the spectral
structure is present from the initial target onset through the saccade.
 }{ 
  \centerline{\ps{5in}{!}{Blank.eps}}
  }{fig:SU.Spectra}

\efig{ {\bf Spectral structure in LFP activity}
Line plots showing the spectrum of LFP activity during the memory and
baseline periods in the preferred and anti-preferred directions.

a) Memory (solid) and baseline (dashed) activity at a single site during
saccades to the preferred direction.
b) Memory activity at a single site during
saccades to the preferred (solid) and anti-preferred (dashed) direction.
c) Population average memory (solid) and baseline (dashed) activity 
during saccades to the preferred direction.
d) Population average memory activity during
saccades to the preferred (solid) and anti-preferred (dashed) direction.

Gamma band activity is elevated in the memory period for the preferred
direction at a single site and in the population average.
The narrow peak in a recording from a single site is broadened in the 
population average.
}{ 
  \centerline{\ps{5in}{!}{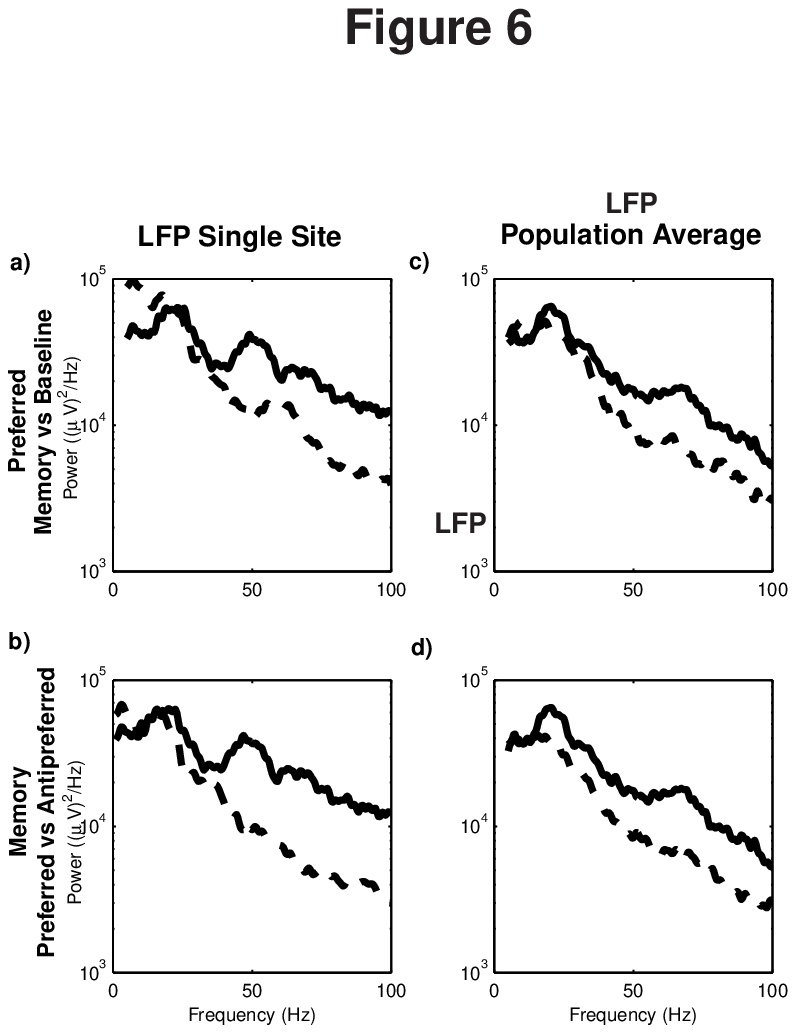}}
  }{fig:LFP.Line.Spectra}

\efig{ {\bf Spectral structure in LFP activity }
Spectrograms of LFP activity averaged across trials
during saccades to the preferred and anti-preferred direction.
Time is on the horizontal axis and frequency is on the vertical axis.
Spectra are color-coded on a log scale.

a) Activity from a single site in the preferred direction.  
b) Activity from a single site in the anti-preferred direction.  
c) Population average activity in the preferred direction from one monkey.  
d) Population average activity in the anti-preferred direction fom one
monkey.  

A strong increase in gamma band
50--90Hz activity localized to the  memory period is present
during saccades to the preferred direction in a single site
and the population average. This is  not seen in
activity in the anti-preferred direction.  
}{ 
  \centerline{\ps{5in}{!}{Blank.eps}}
  }{fig:LFP.Spectra}

\efig{ {\bf SU - LFP coherency}
Coherency between SU and LFP activity
at a single site in the preferred direction.  

Time is on the
horizontal axis and frequency on the vertical axis.
Coherence  is color-coded on a linear scale.  The phase of the
coherency is indicated
by an arrow at each frequency that the coherence is 
significant ($p<0.01$).  The mean firing rate of the SU activity and
rasters showing the trial-by-trial activity are displayed above.  Eye
movement traces and behavioral events are shown below.

Coherence is only significant during the
memory period in the gamma frequency band, 40--90Hz.
 }{ 
  \centerline{\ps{3.4in}{!}{Blank.eps}}
  }{fig:Coherency}

\efig{  {\bf SU - LFP population coherency }
Population average coherency of SU and LFP activity
in the preferred direction
from one monkey.
Time is on the
horizontal axis and frequency on the vertical axis.
The coherence is color-coded on a linear scale. The phase of the
coherency is indicated
by an arrow at each frequency where the coherence 
is significant ($p<0.01$).  Behavioral events are shown below.
The significant increase in single site
coherence during the memory period is also present in the population average.
This indicates that the phase of the coherency is constant across
different recordings in area LIP.
}{ 
  \centerline{\ps{3.4in}{!}{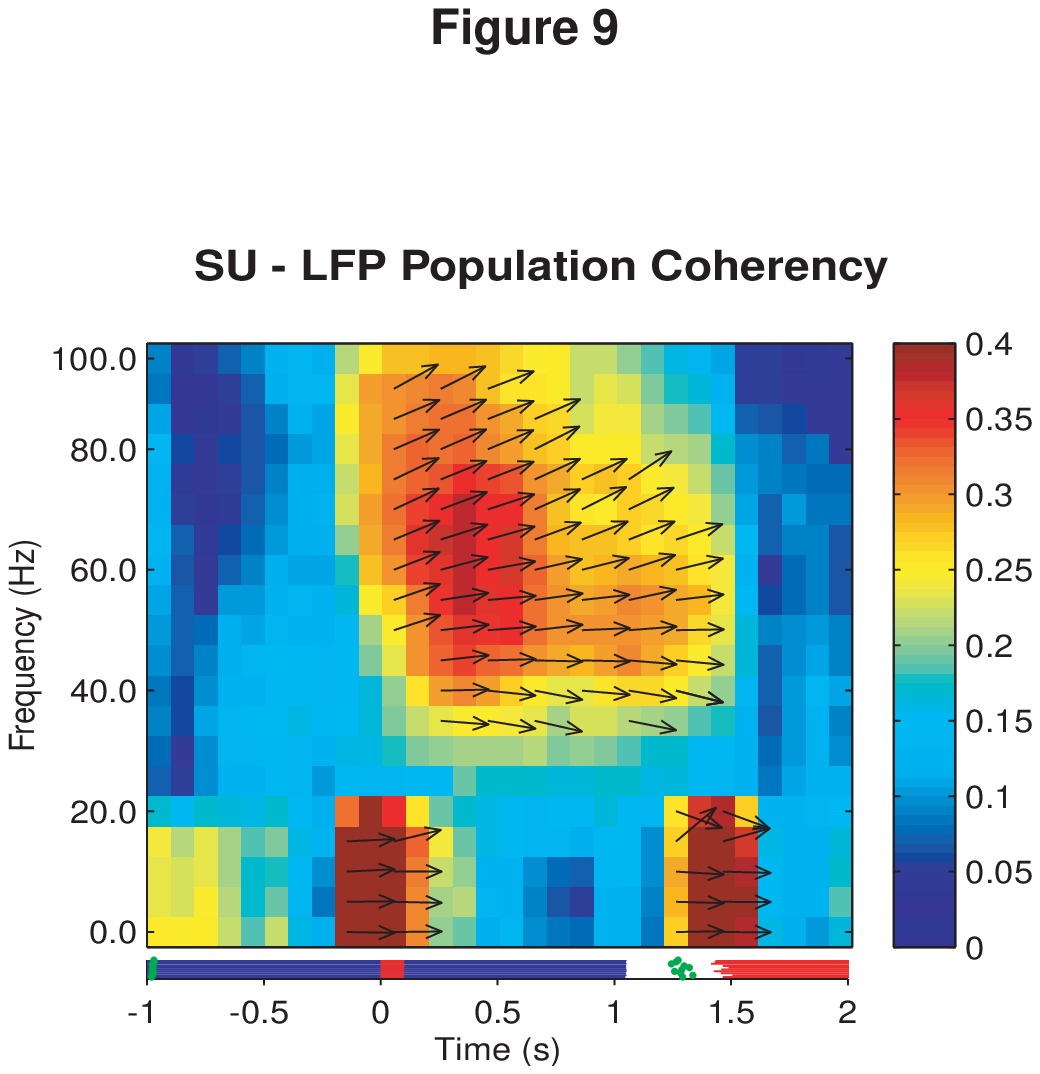}}
  }{fig:PopulationCoherency}

\efig{  {\bf Spike-triggered potential}
Spike-triggered average potential for activity from a single cell
at a single site during trials for
saccades to the preferred direction.  
Baseline activity and
memory period activity were averaged separately.
  
a) Spike-triggered average potential 
for SU activity during the baseline period.  
b) Spike-triggered average potential during the memory period

During the baseline period structure is present before the spike 
indicating weak locking to the extracellular 
potential in the generation of a spike.  Memory period
activity shows oscillatory structure centered on the spike
event time indicating locking to temporal structure in the
extracellular potential.
}{ 
  \centerline{\ps{3.4in}{!}{Blank.eps}}
  }{fig:SpikeTriggeredPotential}

\efig{  {\bf Phase histogram} 
The distribution of the phase of the $70 \pm 20$Hz activity in the 
LFP at the spike event times for all 
cells in the data set during saccades to their
preferred direction.  Memory (solid) and baseline (dashed)
period distributions are compared.  
Strong locking to the $70 \pm 20$Hz field potential oscillation at zero 
phase lag is present during the memory period 
and not during the baseline period.  
}{
  \centerline{\ps{3.4in}{!}{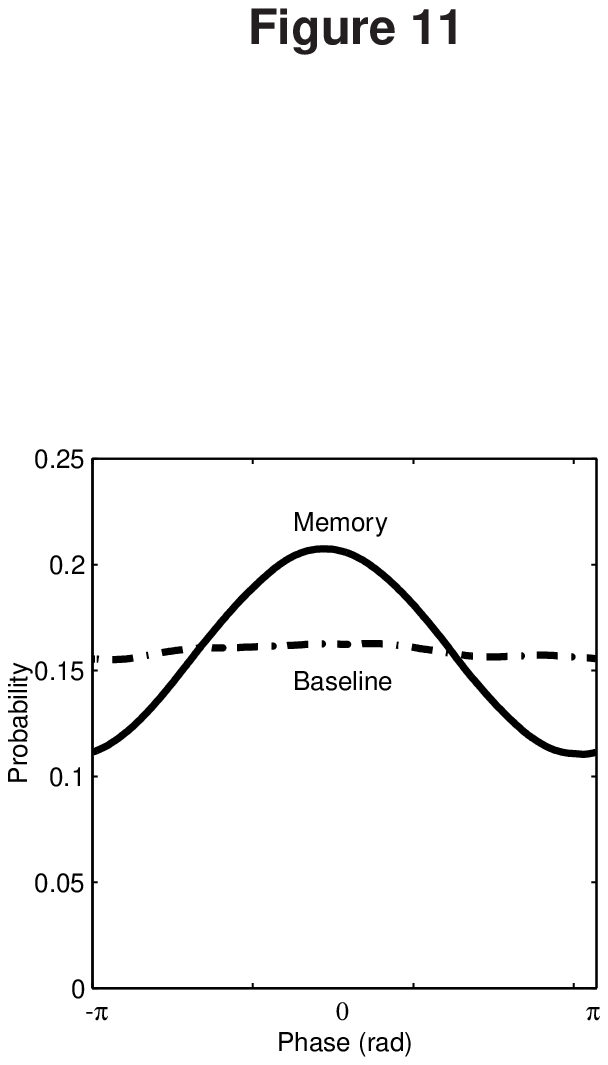}}
  }{fig:PhaseHistogram}

\efig{  {\bf Dynamics of 70Hz power in SU and LFP activity}  
The dynamics of the power of neuronal activity in
the $70 \pm 20$Hz gamma frequency band

a)  LFP (solid) and SU (dashed) gamma band
activity at a from a single cell at a single site.
b)  Population average of LFP (solid) and SU (dashed) activity 
in one monkey.  

Dynamics of SU and LFP gamma band 
activity are very similar in all phases of the trial.
}{ 
  \centerline{\ps{3.4in}{!}{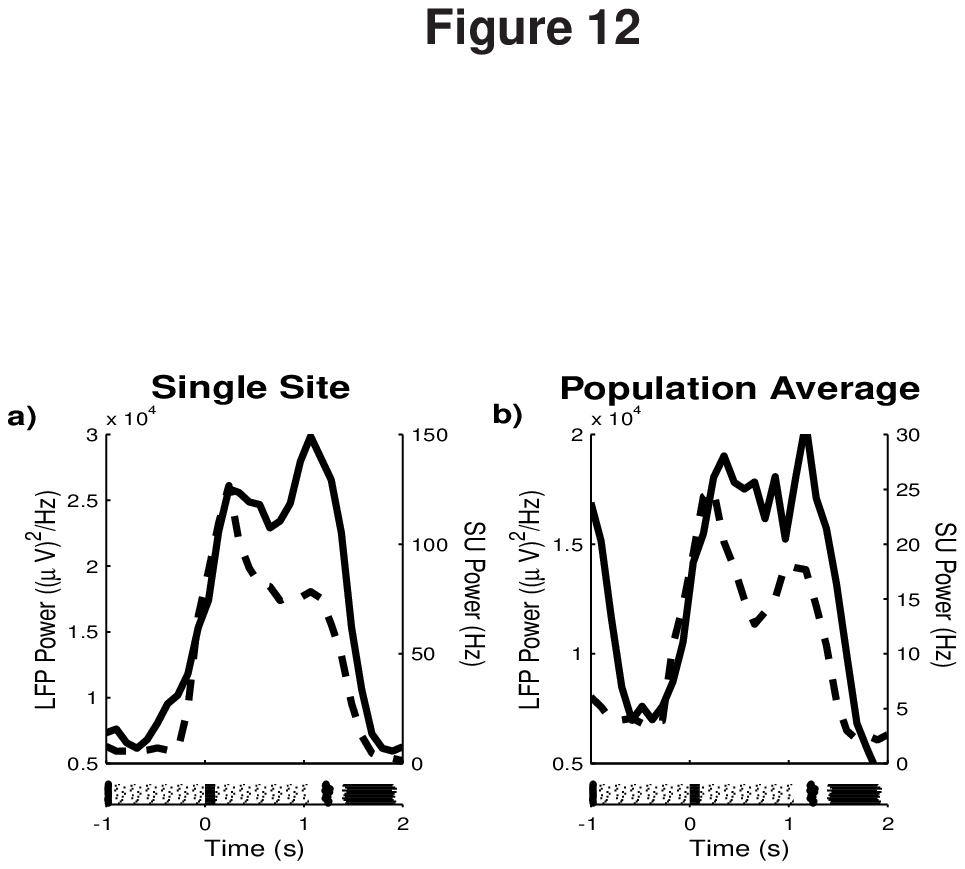}}
  }{fig:Dynamics}

\efig{ {\bf Tuning of 70Hz power in SU and LFP activity}
Tuning of the power of neuronal activity in
the $70 \pm 20$Hz gamma frequency band at a single site.

a) Baseline (dot-dash) memory
(solid) and perisaccadic (dashed) SU activity at 70Hz is
shown against saccade direction.
The preferred direction aligned to the center of the plot.
b)  Baseline (dot-dash) memory
(solid) and perisaccadic (dashed) LFP activity at 70Hz is
shown against saccade direction.
The preferred direction aligned to the center of the plot.

Tuning of SU and LFP gamma band memory period activity are very similar.
Memory and perisaccadic activity are tuned while
the baseline period shows no tuned activity.
 }{ 
  \centerline{\ps{3.4in}{!}{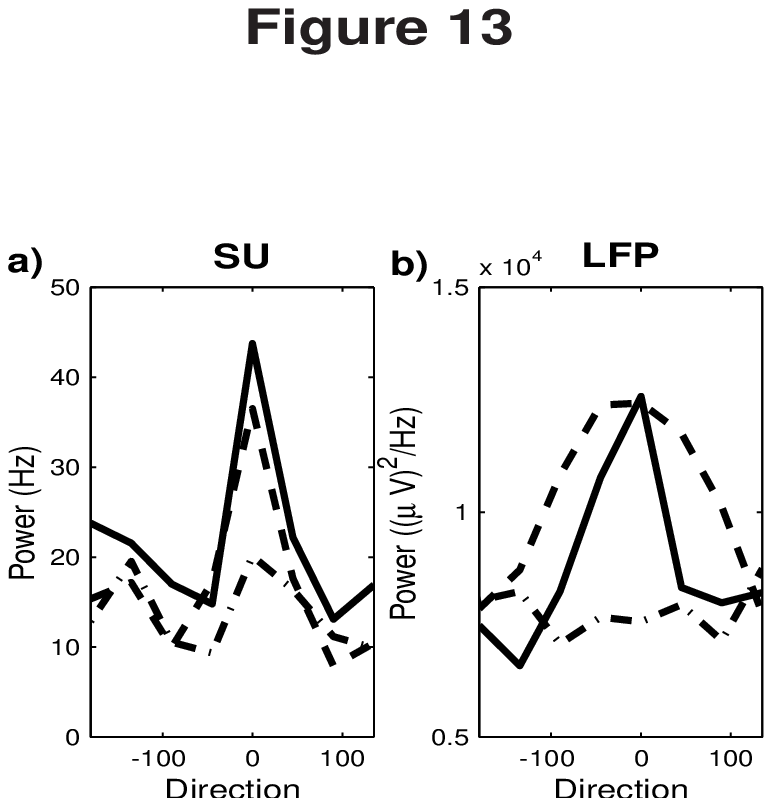}}
  }{fig:Tuning}

\efig{ {\bf SU and LFP 70Hz power}
Power in SU and LFP activity in the $70 \pm 20$Hz
frequency band averaged across trials for saccades to different 
directions.

a) 2d plot of SU activity from a single cell.  Power is color-coded on a linear
scale.  Time is on the horizontal axis.
Saccade direction is on the vertical axis.  The preferred direction is
aligned to 0 degrees.  Behavioral events are shown below.
b) 2d plot of LFP activity from a single site shown in the same way as
SU activity in panel a).

 }{ 
  \centerline{\ps{5in}{!}{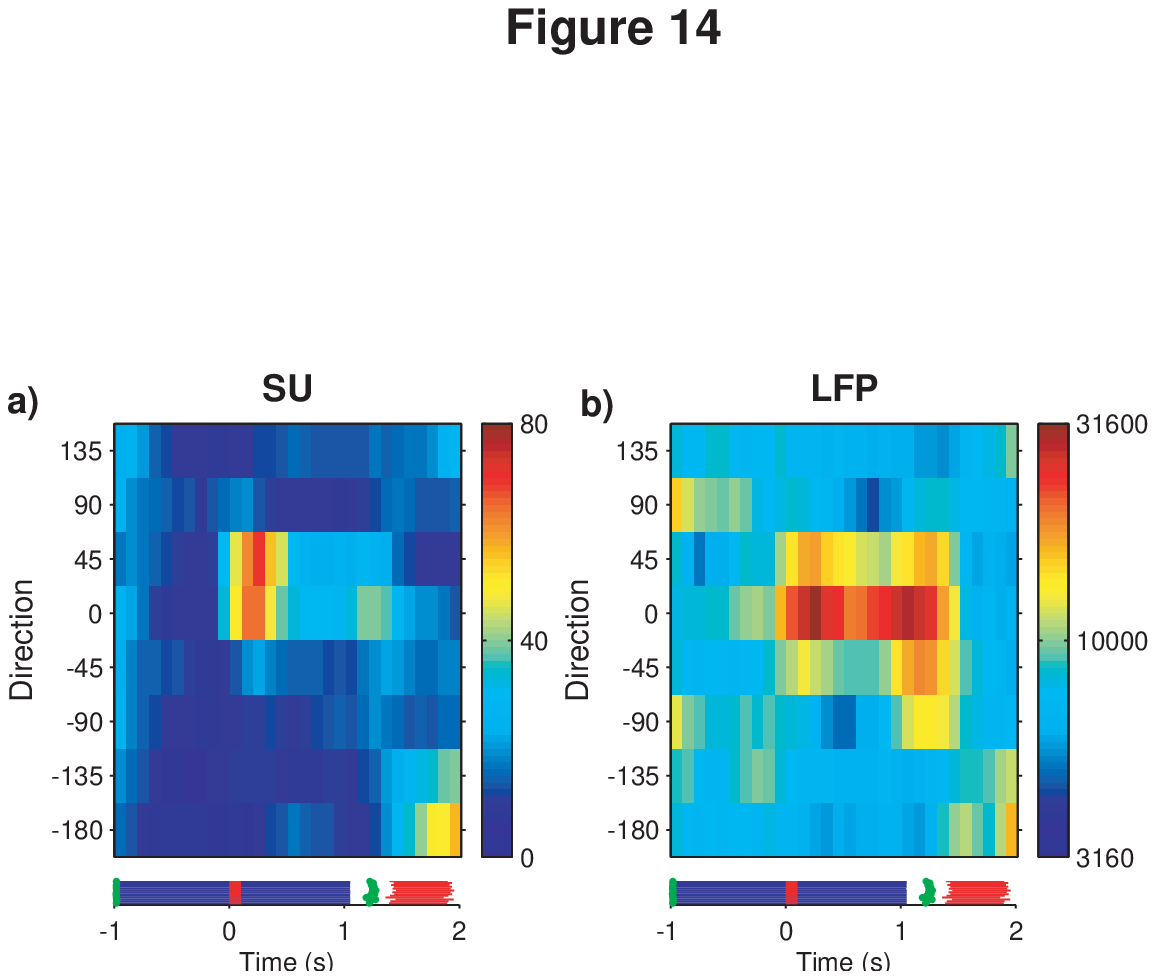}}
  }{fig:LFPvsSU.70Hz}

\efig{ {\bf LFP power at $20$Hz and $70$Hz}
The LFP activity is compared between frequency bands at $70 \pm 20$Hz 
and $20 \pm 5$Hz for trials with saccades to the preferred directions.

a)  LFP activity at 70Hz (solid) and 20Hz (dashed) from a single site.  
b)  Population average LFP activity at 70Hz (solid) and 20Hz
(dashed) from one monkey.  

70Hz activity is elevated 
from the initial target onset through the saccade.  
20Hz activity rises toward the end of
the memory period and is suppressed peri-saccadically.  
The elevated 70Hz activity may be related to movement planning.
The suppression in 
20Hz activity peri-saccadically may be related to movement
execution.  The rise in 20Hz activity toward the end of the memory period 
may reflect preparatory aspects of the task. }{ 
  \centerline{\ps{3.4in}{!}{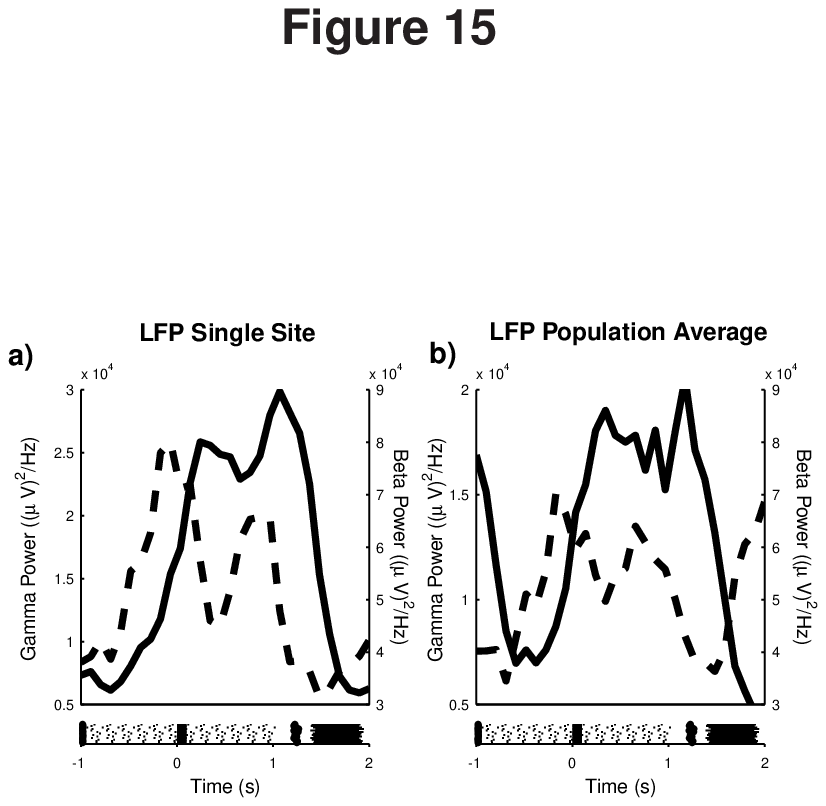}}
  }{fig:LFP.Lo-Hi.Activity}

\end{document}